\documentclass[twocolumn,pra,floatfix,showpacs,superscriptaddress]{revtex4}
\usepackage{amssymb}
\usepackage{amsmath}
\usepackage{graphicx}
\usepackage{float}

\begin{document}
\title{The Group Velocity of a Probe Light in an Ensemble
of $\Lambda$-Atoms under Two-Photon Resonance}
\author{Y. Li}
\affiliation{Institute of Theoretical Physics, Chinese Academy of Sciences,
Beijing, 100080, China}
\author{C. P. Sun}\email{suncp@itp.ac.cn;www.itp.ac.cn/~suncp}
\affiliation{Institute of Theoretical Physics, Chinese Academy of Sciences,
Beijing, 100080, China}


\begin{abstract}
We study the propagation of a probe light in an ensemble of
$\Lambda$-type atoms, utilizing the dynamic symmetry as recently
discovered when the atoms are coupled to a classical control field
and a quantum probe field {[Sun {\it et al.,} Phys. Rev. Lett.
{\bf 91}, 147903 (2003)]}. Under two-photon resonance, we
calculate the group velocity of the probe light with collective
atomic excitations. Our result gives the dependence of the group
velocity on the common one-photon detuning, and can be compared
with the recent experiment (E. E. Mikhailov, Y. V. Rostovtsev, and
G. R. Welch, quant-ph/0309173).
\end{abstract}

 \maketitle

Electromagnetically induced transparency (EIT) \cite{1} has become an active
area of theoretical and experimental research \cite{Brandt,Vanier,Lukin97}.
Since the discovery of EIT, a host of new effects and techniques for
light-matter interaction has occurred; e.g. the propagation of ultra-slow
light pulses \cite{2,3}, the storage of light in atomic vapors \cite{4,5} or
in an "atomic crystal" \cite{Sun01}, the cooling of ground state atoms, and
the giant cross-Kerr non-linearity \cite{8}.

A conventional EIT system consists of a vapor cell with 3-level atoms near
resonantly coupled to two classical fields (from the control and probe
lasers) \cite{1,2,Scullybook}. To investigate its application as a quantum
memory or for transferring quantum information between light (photons) and
atoms, several groups \cite{Lukin00-ent,11,Lukin01,Fleischhauer01} replaced
the classical probe laser field with a weak quantum field. By adiabatically
changing the coupling strength of the classic control field, it was shown
that the propagation of the quantum probe field can be coherently controlled
via the {\it so-called} dark sates and dark-state polaritons. The recent
experiments \cite{4,5} on light storage have further demonstrated the
possibility of using this system for storage of quantum information.

In most studies of quantum memory based on EIT systems
\cite{Sun01,11}, both the probe and control fields are required to
be on resonance with the relevant (one-photon) atomic transitions.
We note, however, on-resonance EIT is in fact not a prerequisite
for achieving significant group velocity reduction
\cite{Lukin-rmp}. More generally, the EIT phenomenon occurs when
the probe and control fields are two-photon Raman resonant with the
$\Lambda $-type atoms. Refs.
\cite{Deng01,Deng02,Greentree,Kocharovskaya} reported theoretical
and experimental results on significant group-velocity reduction
when both fields are classical and two-photon resonant with the
atoms. A more recent experiment \cite{Welch2003} demonstrated the
dependence of ultra-slow group velocity on the probe light detuning
under two-photon resonance, with or without a buffer gas. Some of
their experimental results are, however, difficult to explain using
the conventional EIT theory with a single atom.

In this article, we revisit the above two-photon resonant EIT
system with the dynamic symmetry analysis as developed earlier
\cite{Sun01}. In Ref. \cite{Sun01}, we find the EIT system, which
is consisted of $\Lambda$-type atoms {\it exactly} resonantly
coupled by the quantum probe light and the classical control
light, possesses a hidden dynamic symmetry described by the
semi-direct product of quasi-spin $SU(2)$ and the boson algebra of
the excitons. Here we'll further prove that the same hidden
dynamic symmetry persists in the more general two-photon resonant
case. This observation allows us to build a dynamic equation
describing the propagation of the probe light in this atomic
ensemble with atomic collective excitations. We calculate the
group velocity of the quantum probe field, and investigate how it
depends on the detuning of the control and probe fields. Put aside
the influence of atomic spatial motion, atomic collisions, and
buffer gas atoms, our results are consistent with some of the
recent experiment \cite{Welch2003}.

\begin{figure}[h]
\hspace{48pt}\includegraphics[width=4cm,height=4cm]{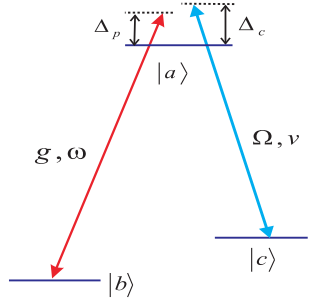}
\caption{A 3-level $\Lambda$-atom coupled to a classic control and a quantum
probe fields with respective detuning $\Delta _{{\rm c}}$ and $\Delta _{{\rm %
p}}$. When $\Delta _{{\rm p}}$=$\Delta_{{\rm c}}$, the system satisfies the
two-photon resonance EIT condition.}
\label{fig1}
\end{figure}

We consider an ensemble of $N$ 3-level $\Lambda$-type atoms, coupled to a
classical control field and a quantum probe field as shown in Fig. \ref{fig1}%
. The atomic levels are labelled as the ground state $|b\rangle $, the
excited state $|a\rangle $ and the final state $|c\rangle $. The atomic
transition $|a\rangle \leftrightarrow |b\rangle $ with energy level
difference $\omega _{{\rm ab}}$ $=\omega _{{\rm a}}-\omega _{{\rm b}}$ is
coupled to the quantum probe field of frequency $\omega $ with the coupling
coefficient $g$ and the detuning $\Delta _{{\rm p}}=\omega -\omega _{{\rm ab}%
}$, while the atomic transition $|a\rangle \leftrightarrow |c\rangle $ with
energy level difference $\omega _{{\rm ac}}$ is driven by a classical
control field of frequency $\nu$ with the Rabi-frequency $\Omega $ and the
detuning $\Delta _{{\rm c}}=\nu-\omega _{{\rm ac}}$.

In the interaction picture, the interaction part of our Hamiltonian reads ($%
\hbar =1$)
\begin{equation}
H_{I}=-\Delta _{{\rm p}}S+(g\sqrt{N}aA^{\dagger }+e^{i(\Delta _{{\rm p}%
}-\Delta _{{\rm c}})t}\Omega T_{+}+h.c.),  \label{hi01}
\end{equation}%
in terms of the collective quasi-spin operators

\begin{equation}
S=\sum_{j=1}^{N}\sigma _{{\rm aa}}^{(j)},\ \ A^{\dagger }=\frac{1}{\sqrt{N}}%
\sum_{j=1}^{N}\sigma _{{\rm ab}}^{(j)},\ \ T_{+}=\sum_{j=1}^{N}\sigma _{{\rm %
ac}}^{(j)}.
\end{equation}%
Here $\sigma_{\mu\nu}^{(j)}=|\mu\rangle_{jj}\langle \nu |$ is the flip
operator of the $j^{th}$ atom from state $|\mu\rangle _{j}$ to $%
|\nu\rangle_{j}$ ($\mu,\nu =a,b,c$); and $a^{\dagger}$ ($a$) is the creation
(annihilation) operator of the probe light. In the large $N$ limit with low
atomic excitations, only a few atoms occupy states $|a\rangle $ or $%
|c\rangle $ \cite{q-defor}, and the atomic collective excitations
of the atoms behave as bosons since in this case they satisfy the
bosonic commutation
relation $[A,A^{\dagger }]=1$. When at two-photon resonance defined by $%
\Delta_{{\rm p}}$=$\Delta_{{\rm c}}$, the Hamiltonian (\ref{hi01}) is
time-independent and thus there exists the same dark state and dark state
polariton as shown before for the one-photon resonant case \cite{Sun01,11}.

We note that the above Hamiltonian is expressed in terms of the collective
dynamic variables $S$, $A$, $A^{\dagger }$, $T_{-}=(T_{+})^{\dagger }$, and $%
T_{+}$. To properly describe both the probe light propagation and
the cooperative motion of the atomic ensemble stimulated by the
two fields, we consider the closed Lie algebra generated by $A$,
$A^{\dagger }$, $T_{-}$ and $T_{+}$. To this end a new pair of
atomic collective excitation operators
\begin{equation}
C=\frac{1}{\sqrt{N}}\sum_{j=1}^{N}\sigma _{{\rm bc}}^{(j)},\ \ C^{\dagger
}=(C)^{\dagger }
\end{equation}%
are introduced here to form a closed algebra. In the low
excitation limit when a few atoms occupy states $|a\rangle $ and
$|c\rangle $, the corresponding atomic collective excitations also
behave as bosons since they satisfy the bosonic commutation
relation $[C,C^{\dagger }]=1$. These atomic collective excitations
are independent of each other in the same limit because of the
vanishing commutation relations $[A,C]=0,[A,C^{\dagger
}]\rightarrow 0$ by a straightforward calculation. Together with
the above commutators the following basic commutation relations
\begin{eqnarray}
\lbrack S,C^{\dagger }] &=&0,[A,S]=A,  \nonumber \\
\lbrack T_{-},C^{\dagger }] &=&0,[T_{+},C^{\dagger }]=A^{\dagger }
\end{eqnarray}%
define a dynamic symmetry hidden in our dressed atomic ensemble
described by the semi-direct-product algebra containing the
algebra $SU(2)$ generated by $T_{-}$ and $T_{+}$.

We now calculate the probe field group velocity from the time-dependent
Hamiltonian (\ref{hi01}). With the help of the above dynamic algebra, we can
write down the Heisenberg equations of operators $A$ and $C$ as
\begin{eqnarray}
\dot {A} &=&-(\Gamma _{{\rm A}}-i\Delta _{{\rm p}})A-ig\sqrt{N}a
\nonumber \\
&&-ie^{i(\Delta _{{\rm p}}-\Delta _{{\rm c}})t}\Omega C+f_{{\rm A}}(t), \\
\dot {C} &=&-\Gamma _{{\rm C}}C-ie^{-i(\Delta _{{\rm p}}-\Delta _{%
{\rm c}})t}\Omega A+f_{{\rm C}}(t),  \nonumber
\end{eqnarray}%
where we have phenomenologically introduced the decay rates $\Gamma _{{\rm A}%
}$ and $\Gamma _{{\rm C}}$ of the states $|a\rangle $ and
$|c\rangle $, and $f_{{\rm A}}(t)$ and $f_{{\rm C}}(t)$ are
the quantum fluctuation of operators with $\left\langle f_{{\rm \alpha }%
}(t)f_{{\rm \alpha }}(t^{\prime })\right\rangle \neq 0$, but $\left\langle
f_{{\rm \alpha }}(t)\right\rangle =0$, $(\alpha =A,C)$.

To find the steady state solution for the above motion equations of atomic
coherent excitation, it is convenient to remove the fast changing factors by
making a transformation $C=\tilde{C}e^{-i(\Delta _{{\rm p}}-\Delta _{{\rm c}%
})t}$. The steady state solution can be achieved{\rm \ }from the transformed
equations
\begin{eqnarray}
\dot {A} &=&-(\Gamma _{{\rm A}}-i\Delta _{{\rm p}})A-ig\sqrt{N}%
a-i\Omega \tilde{C}+f_{{\rm A}}(t),  \nonumber  \label{at} \\
\dot{\tilde{C}} &=&-\Gamma _{{\rm C}}\tilde{C}+i(\Delta _{{\rm p}%
}-\Delta _{{\rm c}})\tilde{C}-i\Omega A+f_{{\rm C}}(t),  \label{ct}
\end{eqnarray}%
by letting $\dot{A}=\dot{\tilde{C}}=0$. The mean expression of $A$
explicitly obtained is
\begin{equation}
\left\langle A\right\rangle =\frac{-ig\sqrt{N}[\Gamma _{{\rm C}}-i(\Delta _{%
{\rm p}}-\Delta _{{\rm c}})]\left\langle a\right\rangle }{(\Gamma _{{\rm A}%
}-i\Delta _{{\rm p}})[\Gamma _{{\rm C}}-i(\Delta _{{\rm p}}-\Delta _{{\rm c}%
})]+\Omega ^{2}}.
\end{equation}

It is noticed that the single-mode probe quantum light is described by
\begin{equation}
E(t)=\varepsilon e^{-i\omega t}+h.c.\equiv \sqrt{\frac{\omega }{2V\epsilon
_{0}}}ae^{-i\omega t}+h.c.,
\end{equation}
where $V$ is the effective mode volume, which for simplicity is
chosen to be equal to the interaction volume. While its
corresponding polarization is
\begin{equation}
\left\langle P\right\rangle =\left\langle p\right\rangle e^{-i\omega
t}+h.c.\equiv \epsilon _{0}\chi \left\langle \varepsilon \right\rangle
e^{-i\omega t}+h.c.,
\end{equation}%
where $\chi =\left\langle p\right\rangle /(\left\langle \varepsilon
\right\rangle \epsilon _{0})$ is the susceptibility. Let $\mu $ denote the
dipole moment between states $\left\vert a\right\rangle $ and $\left\vert
b\right\rangle $. The average polarization

\begin{equation}
\left\langle p\right\rangle =\mu \left\langle \sum_{j=1}^{N}\sigma _{{\rm ba}%
}^{(j)}\right\rangle /V=\frac{\mu \sqrt{N}}{V}\left\langle A\right\rangle
\end{equation}%
can be expressed here in terms of the average of the exciton operators $A$.
Since the coupling coefficient $g=-\mu \sqrt{\frac{\omega }{2V\epsilon _{0}}}
$, the susceptibility can be obtained as
\begin{equation}
\chi =\frac{2ig^{2}N(\Gamma _{{\rm C}}-i(\Delta _{{\rm p}}-\Delta _{{\rm c}%
}))}{\omega \lbrack (\Gamma _{{\rm A}}-i\Delta _{{\rm p}})(\Gamma _{{\rm C}%
}-i(\Delta _{{\rm p}}-\Delta _{{\rm c}}))+\Omega ^{2}]}.
\end{equation}%
The real and imaginary parts $\chi _{1}$ and $\chi _{2}$ of this complex
susceptibility $\chi =\chi _{1}+i\chi _{2}$ can be explicitly expressed as
\begin{eqnarray}
\chi _{1} &=&\frac{[(\Delta _{{\rm p}}-\Delta _{{\rm c}})\Theta -\Gamma _{%
{\rm C}}\Xi ]F}{\Theta ^{2}+\Xi ^{2}},  \label{x1} \\
\chi _{2} &=&\frac{[\Gamma _{{\rm C}}\Theta +(\Delta _{{\rm p}}-\Delta _{%
{\rm c}})\Xi ]F}{\Theta ^{2}+\Xi ^{2}},  \label{x2}
\end{eqnarray}%
where $F=2g^{2}N/\omega $ and
\begin{eqnarray}
\Theta &=&\Gamma _{{\rm A}}\Gamma _{{\rm C}}-\Delta _{{\rm p}}(\Delta _{{\rm %
p}}-\Delta _{{\rm c}})+\Omega ^{2},  \nonumber \\
\Xi &=&\Delta _{{\rm p}}\Gamma _{{\rm A}}+\Gamma _{{\rm A}}(\Delta _{{\rm p}%
}-\Delta _{{\rm c}}).
\end{eqnarray}

%
\begin{figure}[h]
\begin{center}
\includegraphics[width=9cm,height=6.5cm]{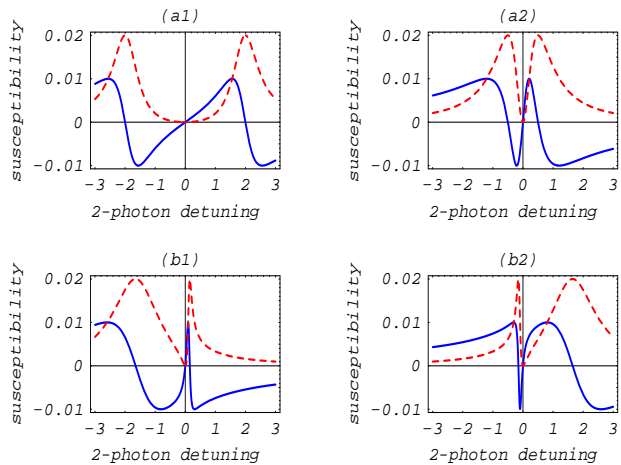}
\end{center}
\caption{ Real part $\protect\chi_1$ (solid) and imaginary part
$\protect\chi _2$ (dashed) of the linear susceptibility vs
two-photon detuning $\Delta $ (=$\Delta_{{\rm p}} -\Delta_{{\rm
c}}$) in normalized units according to: (a1) $\Omega =2$, $\Delta
_{{\rm c}}=0$; (a2) $\Omega =1/2$, $\Delta _{{\rm c}}=0$; (b1,b2)
$\Omega =1/2$, $\Delta _{{\rm c}}=\pm 1.5$. Other parameters are
given as: $\Gamma _{{\rm A}}=1$, $\Gamma _{{\rm C}}=10^{-4}$,
$g\protect\sqrt{N}=100$ and $\protect\omega _{{\rm ab} }=10^{6}$.}
\end{figure}

It is well-known that $\chi _{1}$ and $\chi _{2}$ are related to dispersion
and absorption, respectively. In Fig. 2, $\chi _{1}$ and $\chi _{2}$ are
plotted versus the two-photon detuning $\Delta (=\Delta _{{\rm p}}-\Delta _{%
{\rm c}})$. In Figs. 2(a2,b1,b2), $\Delta _{c}=0,\pm 1.5$
respectively and other parameters are fixed. When $\Delta
\rightarrow 0$, both $\chi _{1}$ and $\chi _{2}$ are almost equal
to zero. This result is
consistent with that in the case of one-photon on-resonance EIT \cite%
{Scullybook,Lukin01}. This fact shows that the medium indeed becomes
transparent when driven by the classical control field as long as the system
is prepared in the two-photon resonance ($\Delta =\Delta _{{\rm p}}-\Delta _{%
{\rm c}}=0$). We also notice that the width of the transparency
window (which is determined by $\chi _{2}$) also depends on the
Rabi frequency $\Omega $. It can be obviously observed from Fig.
2(a1) (where $\Delta _{{\rm c}}=0 $, $\Omega =2$) compared with
Fig. 2(a2) (where $\Delta _{{\rm c}}=0$, $\Omega =1/2$).

Next we consider the properties of refraction and absorption of the
single-mode probe light in the atomic ensemble medium in more detail. To
this end we analyze the complex refractive index
\begin{equation}
n(\omega )=\sqrt{\epsilon (\omega )}=\sqrt{1+\chi },  \label{n}
\end{equation}%
and generally the real and imaginary parts, $n_{1}$ and $n_{2}$, of $n$ are
respectively
\begin{eqnarray}
n_{1} &=&\sqrt{\frac{[(1+\chi _{1})^{2}+\chi _{2}^{2}]^{1/2}+(1+\chi _{1})}{2%
}},  \label{n1} \\
n_{2} &=&\sqrt{\frac{[(1+\chi _{1})^{2}+\chi _{2}^{2}]^{1/2}-(1+\chi _{1})}{2%
}}{\rm sgn}(\chi _{2}),
\end{eqnarray}%
where ${\rm sgn}(\chi _{2})$=$+1$ $(-1)$ if $\chi _{2}>0$ $(<0)$, $n_{1}$
represents the refractive index of the medium and $n_{2}$ is the associated
absorption coefficient. Together with the formulae for the group velocity of
the probe light
\begin{equation}
v_{{\rm g}}(\Delta _{{\rm p}},\Delta _{{\rm c}})=\frac{c}{{\it {Re}[n+\omega
{\rm d}n/{\rm d}\omega ]}}=\frac{c}{n_{1}+\omega \frac{{\rm d}n_{1}}{{\rm d}%
\omega }}  \label{vg1}
\end{equation}%
(where $c$ is the light velocity in vacuum) depending on the frequency
dispersion, one can obtain the explicit expression for the group velocity $%
v_{{\rm g}}$ from Eqs. (\ref{x1}-\ref{n1}) for arbitrary reasonable values
of $\Delta _{{\rm p}}$ and $\Delta _{{\rm c}}$. Now, we consider the group
velocity of the probe light $v_{{\rm g}}$ for the two-photon resonance,
where $\chi _{1}$ and $\chi _{2}$ are almost zero. We find approximately
\[
n_{1}\simeq 1+\chi _{1}/2\rightarrow 1,\ \ n_{2}\simeq \chi _{2}\rightarrow
0,
\]%
and $v_{{\rm g}}$ is given briefly \cite {Dogariu} as:
\begin{eqnarray}
v_{{\rm g}}(\Delta _{{\rm c}}) &=&\frac{c}{n_{1}+\omega \frac{{\rm d}n_{1}}{%
{\rm d}\omega }|_{\Delta _{{\rm p}}=\Delta _{{\rm c}}}}  \nonumber \\
&=&\frac{c}{1+\frac{\omega }{2}\frac{{\rm d}\chi _{1}}{{\rm d}\omega }%
|_{\Delta _{{\rm p}}=\Delta _{{\rm c}}}}.  \label{vg2}
\end{eqnarray}%
It is worth pointing out that, in the calculation of the term $\frac{{\rm d}%
\chi _{1}}{{\rm d}\omega }$, $\Delta _{{\rm p}}$ $(=\omega -\omega _{{\rm ab}%
})$ is a function of $\omega $. In what follows we should make a numerical
calculation of $v_{{\rm g}}(\Delta _{{\rm c}})$ by means of Eqs. (\ref{x1},%
\ref{vg2}) (also or Eqs. (\ref{x1}-\ref{vg1})) since its
analytical expression is too redundant. According to Eq.
(\ref{x1}), the group velocity $v_{{\rm g}}(\Delta _{{\rm c}})$ of
the weak probe field depends on $\Delta _{{\rm c}}$, $\Omega $ and
$g^{2}N$ when given the other relevant parameters (typically,
$\Gamma _{{\rm A}}=1$, $\Gamma _{{\rm C}}=10^{-4}$, $\omega _{{\rm
ab}}=10^{6}$).
%
\begin{figure}[h]
\begin{center}
\includegraphics[width=9cm,height=4cm]{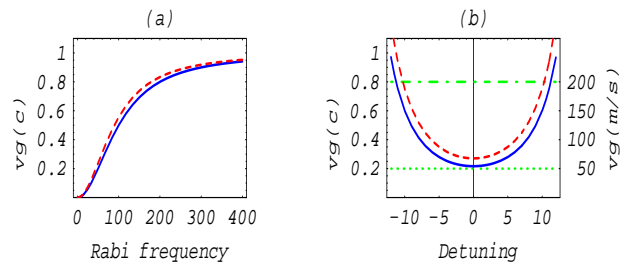}
\end{center}
\caption{The probe light group velocity $v_{{\rm g}}$ vs: (a) Rabi
frequency
$\Omega$ [in normalized units] for $\Delta_{{\rm c}}=5$, and $g%
\protect\sqrt{N} =100$ (blue solid) or $g\protect\sqrt{N} =80$ (red
dashed); (b) the detuning $\Delta _{{\rm c}}$ [in normalized units]
for $\Omega=2g\protect\sqrt{N} =200$ (green dotted-dashed), or
$\Omega=g\protect\sqrt{N}/2 =50$ (green dotted), or $\Omega=0.04 $
and $g\protect\sqrt{N} =100$ (blue solid), or $\Omega=0.04 \Gamma
_{{\rm A}}$ and $g\protect\sqrt{N} =80$ (red dashed), where the
green dotted-dashed and dotted lines are related to the left axis,
and the blue solid and red dashed lines the right axis. Other
parameters are given as: $ \Gamma _{{\rm A}}=1$, $\Gamma _{{\rm
C}}=10^{-4}$ and $\protect\omega _{{\rm ab}}=10^{6}$.}
\end{figure}

Fig. 3(a) shows the dependence of $v_{{\rm g}}(\Delta _{{\rm c}})$
on the Rabi frequency $\Omega $, where the blue solid (red dashed)
line is drawn for $g\sqrt{N}=100$ ($g\sqrt{N}=80$). This provides
one with a technique that can be used to accomplish the storage
and retrieve of the probe pulse. Initially when the probe field
enters into the atomic medium, the Rabi frequency $\Omega $ is
very large (relative to g$\sqrt{N}$) and $v_{{\rm g}}\rightarrow
c$. When one reduces $\Omega $ adiabatically to zero, $v_{{\rm
g}}$ reduces to zero accordingly and then one can store the pulse
in the medium. Conversely, if one wants to retrieve the probe
pulse,
she only needs to increase $\Omega $ adiabatically so as to increase $v_{%
{\rm g}}$. Fig. 3(b) shows the dependence of $v_{{\rm g}}(\Delta _{{\rm c}})$
vs the common detuning $\Delta _{{\rm p}}$(=$\Delta _{{\rm c}}$) under
two-photon resonance EIT. When $\Omega \sim g\sqrt{N}$, $v_{{\rm g}}$ hardly
depends on the detuning $\Delta _{{\rm c}}$ and is close to the simplified
result $\frac{c}{1+g^{2}N/\Omega ^{2}}$ given in Ref. \cite{Lukin01}.
However, when $\Omega \ll g\sqrt{N}$ as denoted by the blue solid and red
dashed curves in Fig. 3(b), $v_{{\rm g}}$ becomes very small and depends on $%
\Delta _{{\rm c}}$. In the symmetric spectral configuration we
find that the group velocity $v_{{\rm g}}$ of the quantum probe
light takes
its minimum near the zero detuning and $v_{{\rm g}}$ increases when $\mid $$%
\Delta _{{\rm p}}$$\mid $ increases in the case of two-photon resonance.
This theoretical result is consistent with the experimental phenomena as
discovered in Ref. \cite{Welch2003} when no buffer gases are used.

Finally we notice that, in our model, the density of the medium is
proportional to the atom number $N$. Fig. 3(a,b) demonstrates how
$v_{g}$ depends on atomic density. In Fig. 3(a) and Fig. 3(b), the
blue solid curve is plotted for a denser medium ($g\sqrt{N}=100$
and $g$ is given as constant) than that of the red dashed curve
($g\sqrt{N}=80$). We also find that a denser medium leads to a
slower $v_{{\rm g}}$, consistent with our physical intuition.

In the present parameters given in this paper, the group velocity
$v_{g}$ is within the zone (0,c). It's remarked that $v_{g}$ can be
negative or superluminal in other $\Lambda$-atoms system as in Ref.
\cite{Godone,Dogariu}. Different from our EIT system, a system
consisted of $\Lambda$-atoms coupled to three optical fields is
studied in \cite{Godone}, and it is the issue of
coherent-population-trapping (not EIT) that is considered in Ref.
\cite{Dogariu}. It's also remarked that a theoretical work
\cite{Kocharovskaya} about the $\Lambda$-atoms EIT system, shows a
negative group velocity can appear since the effect of atomic
spatial motion (or also the buffer gases) in the hot atoms is
considered. Contrarily, the group velocity is always within $(0,c)$
in our EIT system. In our opinion, this difference is mainly due to
our ignoring the effect of the atomic spatial motion and the buffer
gases in this work.

In conclusion, based on the novel algebraic dynamics method, our
theoretical studies on the light propagation in an atomic ensemble
with two-photon
resonance EIT show a similar phenomenon as discovered in the experiment \cite%
{Welch2003}. Our analysis ignores the generated Stokes field, which is also
detected in the above experiment and described in Ref. \cite%
{Stokes01,Stokes02}. We also neglect the influence of atomic
spatial motion, atomic collisions, and the effects of buffer gases,
since in principle these effects can be taken into account as the
perturbations in our present study when the atomic ensemble is
prepared under enough low temperature.

{\it We acknowledge the support of the CNSF (grant No.90203018), the
Knowledge Innovation Program (KIP) of the Chinese Academy of Sciences, and
the National Fundamental Research Program of China (No.001GB309310).}

\end{document}